\documentclass[twocolumn,conference]{IEEEtran}
\usepackage[LGR,T1]{fontenc}
\usepackage[latin9]{inputenc}
\usepackage{float}
\usepackage{mathtools}
\usepackage{amsmath}
\usepackage{amssymb}
\usepackage{graphicx}
\usepackage[unicode=true,
 bookmarks=true,bookmarksnumbered=true,bookmarksopen=true,bookmarksopenlevel=1,
 breaklinks=false,pdfborder={0 0 0},pdfborderstyle={},backref=false,colorlinks=false]
 {hyperref}
\hypersetup{pdftitle={Your Title},
 pdfauthor={Your Name},
 pdfpagelayout=OneColumn, pdfnewwindow=true, pdfstartview=XYZ, plainpages=false}

\makeatletter

\DeclareRobustCommand{\greektext}{%
  \fontencoding{LGR}\selectfont\def\encodingdefault{LGR}}
\DeclareRobustCommand{\textgreek}[1]{\leavevmode{\greektext #1}}
\ProvideTextCommand{\~}{LGR}[1]{\char126#1}

\providecommand{\tabularnewline}{\\}
\floatstyle{ruled}
\newfloat{algorithm}{tbp}{loa}
\providecommand{\algorithmname}{Algorithm}
\floatname{algorithm}{\protect\algorithmname}

\usepackage[caption=false,font=footnotesize]{subfig}

\usepackage{amsfonts,times,amsmath,amssymb,amsthm}
\usepackage{algpseudocode}

\def\app#1#2{%
  \mathrel{%
    \setbox0=\hbox{$#1\sim$}%
    \setbox2=\hbox{%
      \rlap{\hbox{$#1\propto$}}%
      \lower1.1\ht0\box0%
    }%
    \raise0.25\ht2\box2%
  }%
}

\@ifundefined{showcaptionsetup}{}{%
 \PassOptionsToPackage{caption=false}{subfig}}
\usepackage{subfig}
\makeatother

\begin{document}
\title{A Moving Window Based Approach to Multi-scan Multi-Target Tracking}
\author{\IEEEauthorblockN{Diluka Moratuwage\IEEEauthorrefmark{1}, Changbeom Shim \IEEEauthorrefmark{2}, and Yuthika Punchihewa \IEEEauthorrefmark{3}} \newline \IEEEauthorblockA{\IEEEauthorrefmark{1}\IEEEauthorrefmark{2}\IEEEauthorrefmark{3}\textit{School of Electrical Engineering, Computing and Mathematical Sciences, Curtin University, Australia.}} \newline \IEEEauthorrefmark{1} diluka.moratuwage@curtin.edu.au, \IEEEauthorrefmark{2} changbeom.shim@curtin.edu.au, \IEEEauthorrefmark{3} yuthika.gardiyaw@curtin.edu.au.     {\let\thefootnote\relax\footnotetext{This work is supported by the Australian Research Council under Linkage project LP200301507}}}
\maketitle
\begin{abstract}
Multi-target state estimation refers to estimating the number of targets
and their trajectories in a surveillance area using measurements contaminated
with noise and clutter. In the Bayesian paradigm, the most common
approach to multi-target estimation is by recursively propagating
the multi-target filtering density, updating it with current measurements
set at each timestep. In comparison, multi-target smoothing uses all
measurements up to current timestep and recursively propagates the
entire history of multi-target state using the multi-target posterior
density. The recent Generalized Labeled Multi-Bernoulli (GLMB) smoother
is an analytic recursion that propagate the labeled multi-object posterior
by recursively updating labels to measurement association maps from
the beginning to current timestep. In this paper, we propose a moving
window based solution for multi-target tracking using the GLMB smoother,
so that only those association maps in a window (consisting of latest
maps) get updated, resulting in an efficient approximate solution
suitable for practical implementations.
\end{abstract}

\section{Introduction}

In the Bayesian approach to single-target tracking, the current state
(represented by a vector) of the target conditioned on the history
of measurements is recursively propagated in time via the filtering
density \cite{Bar-Shalom2011,Blackman1999}. On the other hand, in
the multi-scan approach (to single-target tracking) the entire history
of the single-target state (conditioned on the history of measurements)
is recursively propagated via the smoothing density \cite{Vo12-smoothing}.
Analogously, in the multi-target state estimation, multi-target state
is propagated using the multi-object filtering density \cite{Vo2013}
and multi-object smoothing density \cite{Vo2019b}, where the multi-target
state and observations are represented as Random Finite Sets (RFSs)
\cite{Mahler2007a}\cite{Mahler2014}.

The Generalized Labeled Multi-Bernoulli (GLMB) filter \cite{Vo2013,VoVoP14}
is an analytic solution to the multi-object filtering density that
can be used to estimate multi-target state with target labels (identities).
An efficient implementation of the GLMB filter has been proposed in
\cite{Vo2017} by sampling its components using a Gibbs sampler \cite{Casella2004}.
This approach was demonstrated to track over a millions targets from
about a billion detections in relatively challenging signal scenarios
\cite{Beard18-largescale}. Due to its versatility and efficiency,
the GLMB filter has been extended to track before detect \cite{PapiVoVoetal15,LiYiHoseinnezhadetal18},
distributed tracking \cite{Fantacci18,LiYi18}, tracking with lineage
\cite{Nguyenetal-CellSpawning21,Bryantetal18}, as well as applied
to computer vision \cite{KimVoVo19,Gostar-Interacting-19,Ong-TPAMI-20},
Doppler tracking \cite{DoNguyen-Sensors-19,ZhuDoppler22}, field robotics
\cite{MoratuwageSLAM-18,MoratuwageSLAM-Sensor-19}, space situational
awareness \cite{WeiNener-18,WeiNener-19,Gaebler-BirthR-20,Gaebler-IDman-20},
multi-object sensor scheduling and path planning \cite{Beard17-CS,Nguyenetal-PP-19,Cai-SSA-19,Nguyen-MMMP-AAAI-20,ZhuPOMDP22}.

A recent research avenue is GLMB smoothing \cite{Vo2019b}, where
Gibbs sampling was used to solve multi-dimensional assignment problem
in multi-scan multi-object tracking. In addition, this Gibbs sampler
was combined with importance to solve the multi-dimensional assignment
problem in multi-sensor multi-object tracking \cite{VoVoBeard19}.
The labeled multi-object states caters for trajectories, and smoothing
improves the multi-object states in the all past timesteps as the
smoothing density is updated with future information. However, even
with single-target smoothing, the computation cost at each timestep
increases, and may result in an unaffordable computational cost for
practical implementation. In the multi-target state estimation, smoothing
poses and even more difficult NP-hard multi-dimensional assignment
problem that keeps increasing its complexity with every timestep.
Therefore, it is desirable to come up with a multi-target smoothing
approach with an approximately similar computational cost per each
update step. 

In this paper, we propose a novel technique to propagate the multi-object
smoothing density with similar computational cost at each timestep.
Instead of updating the entire history of the multi-object state trajectory,
we only update the latest $N$ scans of the multi-object history.
Since the target trajectories in the GLMB smoother are characterised
by association maps (see Section \ref{sec:Background}), we propose
to link the trajectories in the posterior using a moving window based
method. 

Section II summarizes the background information related to multi-object
state estimation, Section III explains the implementation details
of the proposed moving window based multi-target tracker and presents
the pseudo code for the moving window based smoother. Section IV presents
the numerical results, and Section V concludes the paper.

\section{Background\label{sec:Background}}

We follow the same convention as in \cite{Vo2019b}, and summarize
the symbols, notations and definitions in this section. For a given
set $S$, $\mathcal{F}(S)$ denotes the class of finite subsets of
$S$, $1_{S}(\cdot)$ denotes the indicator function, $\delta_{X}[Y]$
denotes Kroneker-$\delta$ function, where $\delta_{X}[Y]=1$ if $X=Y$,
0 otherwise, and $\langle f,g\rangle$ denotes the inner product,
i.e., $\int f(x)g(x)dx$, of two functions $f$ and $g$. The list
of variables $X_{m},X_{m+1},...,X_{n}$ is abbreviated as $X_{m:n}$,
the cardinality of a finite set $X$ is denoted as $|X|$, and for
some function $f$, the product $\prod_{x\in X}f(x)$ is denoted by
the (single scan) multi-object exponential $f^{X}$, with $f^{\emptyset}=1$.

\subsection{Mulit-object States and Trajectories}

Let $\mathbb{X}$ denote the single object state space, $\mathbb{L}_{k}$
denote the space of all labels at time $k$, and $\mathbb{B}_{k}$
denote the space of all birth labels at time $k$. Then, the label
space for all objects up to time $k$ is given by $\mathbb{L}_{k}=\biguplus\nolimits _{s=0}^{k}\mathbb{B}_{s}$
(note that $\mathbb{L}_{k}=\mathbb{L}_{k-1}\uplus\mathbb{B}_{k}$),
and a $\emph{labeled state}$ of an object existing at time $k$ is
given by $\boldsymbol{x}=(x,\ell)\in\mathbb{X}\mathcal{\times}\mathbb{L}_{k}$,
where the vector $x\in\mathbb{X}$ is its kinematic state and $\ell=(s,\iota)$
is a unique label, $s$ is the $\textit{time of birth}$, and $\iota$
is a unique index to distinguish objects born at the same time. A
sequence of labeled states at consecutive times ($s$ to $t$) 
\begin{equation}
\mathbf{\tau}=[(x_{s},\ell),(x_{s+1},\ell),...,(x_{t},\ell)],\label{eq:trajectory}
\end{equation}
with the common label $\ell$ and states $x_{s},x_{s+1},...,x_{t}\in\mathbb{X}$
is called a $\emph{trajectory}$. 

A $\emph{labeled multi-object}$ state at time $i$ is a finite subset
$\boldsymbol{X}_{i}$ of $\mathbb{X}\mathcal{\times}\mathbb{L}_{i}$
with distinct labels. Let $\mathcal{L\,}$:$\mathcal{\,}\mathbb{X}\mathcal{\times}\mathbb{L}_{i}\!\rightarrow\!\mathbb{L}_{i}$
is the projection defined by $\mathcal{L}((x,\ell))\!=\ell$, and
$\mathcal{L}(\boldsymbol{X}_{i})$ is the set of lables of $\boldsymbol{X}_{i}$.
Note that, a valid $\boldsymbol{X}_{i}$ has distinct labels and results
in the $\emph{distinct label indicator}$ $\Delta(\boldsymbol{X}_{i})\!\triangleq\delta_{|\boldsymbol{X}_{i}|}[|\mathcal{L(}\boldsymbol{X}_{i})|]$
to be equals to one. The $\emph{labeled multi-object}$ state can
also be written as $\boldsymbol{X}_{i}=\left\{ \mathbf{\tau(}i\mathbf{)\mathcal{\,}}\text{:}\mathbf{\mathcal{\,}\tau\in}S\right\} $,
where $S$ is a set of trajectories defined to have kinematic states
and distinct labels at each timestep, and $\mathbf{\tau(}i\mathbf{)}$
denotes the labeled state of trajectory $\mathbf{\tau}$ at time $i$.
The trajectory of the object with label $\ell\in\cup_{i=j}^{k}\mathcal{L}(\boldsymbol{X}_{i}\mathbf{)}$
in a given sequence $\boldsymbol{X}_{j:k}$ of labeled multi-object
states in the interval $\{j\mathcal{\,}$:$\mathcal{\,}k\}$ is given
by: 
\begin{equation}
\boldsymbol{x}_{s(\mathbf{\ell}):t(\mathbf{\ell})}^{(\mathbf{\ell})}=[(x_{s(\mathbf{\ell})}^{(\mathbf{\ell})},\mathbf{\ell}),...,(x_{t(\mathbf{\ell})}^{(\mathbf{\ell})},\mathbf{\ell})],\label{eq:trajectory2}
\end{equation}
where $s(\mathbf{\ell})$ and $t(\mathbf{\ell})$ are respectively
the earliest and latest times the label $\mathbf{\ell}$ exists on
the interval $\{j\mathcal{\,}$:$\mathcal{\,}k\}$, and $\boldsymbol{x}_{i}^{(\mathbf{\ell})}=(x_{i}^{(\mathbf{\ell})},\mathbf{\ell})$
denotes the element of $\boldsymbol{X}_{i}$ with label $\ell\in\mathcal{L}(\boldsymbol{X}_{i}\mathbf{)}$
with unlabeled state $x_{i}^{(\mathbf{\ell})}$. The sequnce $\boldsymbol{X}_{j:k}$
can thus be equivalently represented by the set of all trajectories
\begin{equation}
\boldsymbol{X}_{j:k}\equiv\left\{ \boldsymbol{x}_{s(\mathbf{\ell}):t(\mathbf{\ell})}^{(\mathbf{\ell})}:\ell\in\bigcup\nolimits _{i=j}^{k}\mathcal{L}(\boldsymbol{X}_{i}\mathbf{)}\right\} ,\label{eq:set_of_trajectories}
\end{equation}
of all labels in $\cup_{i=j}^{k}\mathcal{L}(\boldsymbol{X}_{i})$.
Furthermore, for any function $h:\uplus_{I\subseteq\{j:k\}}\mathbb{T}_{I}\rightarrow[0,\infty)$,
the multi-scan multi-object exponential \cite{Vo2019b} is defined
as 
\[
[h]^{\boldsymbol{X}_{j:k}}\triangleq[h]^{\{\boldsymbol{x}_{s(\ell):t(\ell)}^{(\ell)}:\ell\in\mathcal{L}(\boldsymbol{X}_{j:k})\}}=\underset{\ell\in\mathcal{L}(\boldsymbol{X}_{j:k})}{\prod}h(\boldsymbol{x}_{s(\ell):t(\ell)}^{(\ell)}),
\]
where for any non-negative integer $n$ and $i_{1}<i_{2}<...<i_{n}$,
$\mathbb{T}_{\{i_{1},i_{2},...,i_{n}\}}\triangleq\big(\mathbb{X}\times\mathbb{L}_{i_{1}}\big)\times....\times\big(\mathbb{X}\times\mathbb{L}_{i_{n}}\big)$,
with $\mathbb{T}_{\emptyset}=\emptyset$. If $j=k$, the multi-scan
exponential reduces to single-scan multi-object exponential $h^{\boldsymbol{X}_{j}}$
defined above.

\subsection{Multi-object System Model}

Given the multi-object state $\boldsymbol{X}_{k-1}$ at time $k-1$,
each object with state $\boldsymbol{x}_{k-1}=(x_{k-1},\ell_{k-1})\in\boldsymbol{X}_{k-1}$
either survies with probability $P_{S,k-1}(\boldsymbol{x}_{k-1})$
and moves to a new state $\boldsymbol{x}_{k}=(x_{k},\ell_{k})$ with
transition density $f_{S,k|k-1}(x_{k}|x_{k-1},\ell_{k})\delta_{\ell_{k-1}}[\ell_{k}]$,
or dies with probability $Q_{S,k-1}(\boldsymbol{x}_{k-1})=1-P_{S,k-1}(\boldsymbol{x}_{k-1})$
at time $k$. Further, each object with label $\ell_{k}$ in birth
label space $\mathbb{B}_{k}$ is either born with probability $P_{B,k}(\ell_{k})$
and state $x_{k}$ with probability density $f_{B,k}(x_{k},\ell_{k})$,
or not born with probability $Q_{B,k}(\ell_{k})=1-P_{B,k}(\ell_{k})$
at time $k$. Thus, the multi-object state $\boldsymbol{X}_{k}$ at
time $k$ is the superposition of the suviving states and new birth
states, and in the standard multi-object dynamic model, the birth
and survival sets are independent of each other, and each object moves
and dies independently of each other. The multi-object transition
density $\boldsymbol{f}_{k|k-1}(\boldsymbol{X}_{k}|\boldsymbol{X}_{k-1})$
captures the multi-object dynamic model, and see \cite{Vo2019b,Vo2013}
its detailed expressions.

\subsection{Multi-object Observation Model}

Let $\boldsymbol{X}_{k}$ and $Z_{k}$ be the multi-object state at
time $k$, and the set of measurements captured by the sensor. Each
object $\boldsymbol{x}\in\boldsymbol{X}_{k}$ either generates a measurement
$z\in Z_{k}$ (with detection probability $P_{D}(\boldsymbol{x})$)
on the measurement space $\mathbb{Z}$ with likelihood $g_{k}(z|\boldsymbol{x})$
or miss-detected (with probability $Q_{D}(\boldsymbol{x})=1-P_{D}(\boldsymbol{x})$).
Additionally, the sensor also produces measurement clutter, which
is modeled by a Poisson RFS with intensity function $\kappa_{k}$
on $\mathbb{Z}$. Conditional on $\boldsymbol{X}_{k}$, the detections
and measurement clutter are independent, and therefore multi-object
observation $Z_{k}$ is the superposition of them.

A map of the form $\gamma_{k}:\mathbb{L}_{k}\rightarrow\{-1:|Z_{k}|\}$
is called an \textit{association map} if it is positive 1-1 (i.e.,
no two distinct labels are mapped to the same positive value). If
$\ell$ generates the $\gamma_{k}(\ell)$-th measurement $\gamma_{k}(\ell)>0$,
if $\ell$ is misdetected $\gamma_{k}(\ell)=0$, and if $\ell$ does
not exist $\gamma_{k}(\ell)=-1$. Then, the multi-object likelihood
function is 
\begin{align}
g_{k}(Z_{k}| & \boldsymbol{X}_{k})\propto\underset{\gamma_{k}\in\Gamma_{k}}{\sum}\delta_{\mathcal{L}(\gamma_{k})}[\mathcal{L}(\boldsymbol{X}_{k})][\psi_{k,Z_{k}}^{(\gamma_{k}\circ\mathcal{L}(\cdot))}(\cdot)]^{\boldsymbol{X}_{k}},\label{eq:likelihood_func_sensor_v}
\end{align}
where $\mathcal{L}(\gamma_{k})\triangleq\{\ell\in\mathbb{L}_{k}:\gamma_{k}(\ell)\geq0\}$
is the set of \textit{live labels} of $\gamma_{k}$, and $\Gamma_{k}$
is the space of all association maps, $\gamma_{k}\circ\mathcal{L}(\cdot)=\gamma_{k}(\mathcal{L}(\cdot))$
and 
\begin{equation}
\psi_{k,\{z_{1:m}\}}^{(i)}(\boldsymbol{x})=\begin{cases}
\frac{P_{D}(\boldsymbol{x})g_{k}(z_{i}|\boldsymbol{x})}{\kappa_{k}(z_{i})} & i>0\\
Q_{D,k}(\boldsymbol{x}) & i=0
\end{cases}.\label{eq:likelihood_update_func_sensor_v_2}
\end{equation}

\subsection{Trajectory Posterior of a Single Object}

The GLMB posterior is written in terms of single object periors, and
the corresponding association weights \cite{Vo2019b}. Thus, it is
informative to take a close look at the trajectory posterior and the
association weight of an object with label $\ell\in\mathbb{L}_{k}$,
before the presenting the GLMB posterior. Recall that $s(\ell)$ and
$t(\ell)$ respectively denote the earliest and latest times on $\{0:k\}$
such that $\ell$ exists. Assuming that $\ell$ generates the sequence
of measurement indices $\alpha_{s(\ell):k}$, its trajectory posterior
at time $k$ can be in one of four possible stages: (i) new born,
$s(\ell)=k$; (ii) surviving, $t(\ell)=k>s(\ell)$; (iii) die at time
$k$, $t(\ell)=k-1$; (iv) died before time $k$, $t(\ell)<k-1$.
Thus, its trajectory posterior at time $k$ is given by, \allowdisplaybreaks
\begin{align}
 & \tau{}_{0:k}^{(\alpha_{s(\ell):k})}(x_{s(\ell):t(\ell)},\ell)\label{eq:paramd_multi_scan_multi_sensor_trajectory_update}\\
 & =\begin{cases}
\negthinspace\frac{\Lambda_{B,k}^{(\alpha_{k})}(x_{k},\ell)}{\bar{\Lambda}_{B,k}^{(\alpha_{k})}(\ell)}, & \negthinspace\negthinspace\negthinspace\negthinspace\negthinspace s(\ell)=k\\
\negthinspace\frac{\Lambda_{S,k|k-1}^{(\alpha_{k})}(x_{k}|x_{k-1},\ell)\tau_{0:k-1}^{(\alpha_{s(\ell):k-1})}(x_{s(\ell):k-1},\ell)}{\bar{\Lambda}_{S,k|k-1}^{(\alpha_{s(\ell):k})}(\ell)}, & \negthinspace\negthinspace\negthinspace\negthinspace\negthinspace t(\ell)=k>s(\ell)\\
\negthinspace\frac{Q_{S,k-1}(x_{k-1},\ell)\tau_{0:k-1}^{(\alpha_{s(\ell):k-1})}(x_{s(\ell):k-1},\ell)}{\bar{Q}_{S,k-1}^{(\alpha_{s(\ell):k-1})}(\ell)}, & \negthinspace\negthinspace\negthinspace\negthinspace\negthinspace t(\ell)=k-1\\
\negthinspace\tau_{0:t(\ell)}^{(\alpha_{s(\ell):t(\ell)})}(x_{s(\ell):t(\ell)},\ell), & \negthinspace\negthinspace\negthinspace\negthinspace\negthinspace t(\ell)<k-1
\end{cases},\nonumber 
\end{align}
where $\tau_{0:k}^{(\alpha_{s(\ell):k-1})}(x_{s(\ell):k-1},\ell)$
is the trajectory posterior at time $k-1$, \allowdisplaybreaks
\begin{align}
\Lambda_{B,k}^{(\alpha_{k})}(x,\ell)= & \ \psi_{k,Z_{k}}^{(\alpha_{k})}(x,\ell)P_{B,k}(\ell)f_{B,k}(x,\ell),\label{eq:paramd_multi_scan_multi_sensor_weight_update_2}\\
\bar{\Lambda}_{B,k}^{(\alpha_{k})}(\ell)= & \int\Lambda_{B,k}^{(\alpha_{k})}(x,\ell)dx,\label{eq:paramd_multi_scan_multi_sensor_weight_update_3}\\
\Lambda{}_{S,k|k-1}^{(\alpha_{k})}(x_{k}|x_{k-1},\ell)= & \ \psi_{k,Z_{k}}^{(\alpha_{k})}(x_{k},\ell)P_{S,k-1}(x_{k-1},\ell)\label{eq:paramd_multi_scan_multi_sensor_weight_update_4}\\
\times & \ f_{S,k|k-1}(x_{k}|x_{k-1},\ell),\nonumber \\
\bar{\Lambda}{}_{S,k|k-1}^{(\alpha_{s(\ell):k})}(\ell)= & \int\tau_{0:k-1}^{(\alpha_{s(\ell):k-1})}(x_{s(\ell):k-1},\ell)\label{eq:paramd_multi_scan_multi_sensor_weight_update_5}\\
\times & \ \Lambda_{S,k|k-1}^{(\alpha_{k})}(x_{k}|x_{k-1},\ell)dx_{s(\ell):k},\nonumber \\
\bar{Q}{}_{S,k-1}^{(\alpha_{s(\ell):k-1})}(\ell)= & \int\tau_{0:k-1}^{(\alpha_{s(\ell):k-1})}(x_{s(\ell):k-1},\ell)\label{eq:paramd_multi_scan_multi_sensor_weight_update_6}\\
\times & \ Q_{S,k-1}(x_{k-1},\ell)dx_{s(\ell):k-1}.\nonumber 
\end{align}
The association weight of $\ell$ is given by \allowdisplaybreaks
\begin{equation}
\eta_{k|k-1}^{(\alpha_{s(\ell):k})}(\ell)=\begin{cases}
\bar{\Lambda}_{B,k}^{(\alpha_{k})}(\ell), & \negthinspace\negthinspace\negthinspace\negthinspace s(\ell)=k\\
\bar{\Lambda}_{S,k|k-1}^{(\alpha_{s(\ell):k})}(\ell), & \negthinspace\negthinspace\negthinspace\negthinspace t(\ell)=k>s(\ell)\\
\bar{Q}_{S,k-1}^{(\alpha_{s(\ell):k-1})}(\ell), & \negthinspace\negthinspace\negthinspace\negthinspace t(\ell)=k-1\\
Q_{B,k}(\ell), & \negthinspace\negthinspace\negthinspace\negthinspace\ell\in\mathbb{B}_{k},\alpha_{k}=-1{}^{V}
\end{cases}\negthinspace\negthinspace.\label{eq:paramd_multi_scan_multi_sensor_weight_update_1}
\end{equation}

\subsection{Multi-object Bayes Recursion}

Given the observation history $Z_{1:k}$, the multi-object posterior
$\boldsymbol{\pi}_{0:k}(\boldsymbol{X}_{\negthinspace0:k})\triangleq\boldsymbol{\pi}_{0:k}(\boldsymbol{X}_{\negthinspace0:k}|Z_{1:k})$
captures all information about the set of objects in the surveillance
region in the interval \{$0:k\}$. It can be written in the recursive
form
\begin{equation}
\boldsymbol{\pi}_{0:k}(\boldsymbol{X}_{\negthinspace0:k})\propto g_{k}(Z_{k}|\boldsymbol{X}_{\negthinspace k})\boldsymbol{f}_{k|k-1}(\boldsymbol{X}_{\negthinspace k}|\boldsymbol{X}_{\negthinspace k-1})\boldsymbol{\pi}_{0:k-1}(\boldsymbol{X}_{\negthinspace0:k-1}).\label{eq:multi_object_bayes_recursion}
\end{equation}
The GLMB smoother \cite{Vo2019b} was proposed to as an analytic solution
to the multi-object posterior recursion. Assuming that there are no
live objects at the beginning, i.e., $\boldsymbol{\pi}_{0}(\boldsymbol{X}_{0})=\delta_{0}[\mathcal{L}(\boldsymbol{X}_{0})]$
with weight $w_{0}^{(\gamma_{0})}=1$, the GLMB posterior at time
$k$ is given by \cite{Vo2019b}

\begin{align}
 & \boldsymbol{\pi}_{0:k}(\boldsymbol{\boldsymbol{X}}_{0:}{}_{k})\propto\label{eq:paramd_multi_scan_multi_sensor_posterior}\\
 & \ \ \ \ \Delta(\boldsymbol{X}_{0:k})\underset{\gamma_{0:k}}{\sum}w_{0:k}^{(\gamma_{0:k})}\delta_{\mathcal{L}(\gamma_{0:k})}[\mathcal{L}(\boldsymbol{X}_{0:k})][\tau_{0:k}^{(\gamma_{0:k}\circ\mathcal{L}(\cdot))}(\cdot)]^{\boldsymbol{X}_{0:k}},\nonumber 
\end{align}
where $\Delta(\boldsymbol{X}_{0:k})\triangleq\prod_{i=0}^{k}\Delta(\boldsymbol{X}_{i})$,
and 
\begin{align}
\negthinspace\negthinspace\negthinspace\negthinspace w_{0:k}^{(\gamma_{0:k})}= & \prod_{j=1}^{k}w_{j}^{(\gamma_{0:j})},\label{eq:paramd_full_gibbs_weight}\\
\negthinspace\negthinspace\negthinspace\negthinspace w_{j}^{(\gamma_{0:j})}= & 1_{\mathcal{F}(\mathbb{B}_{j}\uplus\mathcal{L}(\gamma_{j-1}))}(\mathcal{L}(\gamma_{j}))[\eta_{j|j-1}^{(\gamma_{0:j}(\cdot))}(\cdot)]^{\mathbb{B}_{j}\uplus\mathcal{L}(\gamma_{j-1})}.\label{eq:paramd_factor_gibbs_weight}
\end{align}
It is clear that the GLMB posterior is completely parameterized by
the set of components $\{(w_{0:k}^{(\gamma_{0:k})},\tau_{0:k}^{(\gamma_{0:k})})\}$
indexed by $\gamma_{0:k}$. As per (\ref{eq:multi_object_bayes_recursion})
the number of such components grow exponentially after each measurement
update step, and to achieve tractability, truncation is performed
and retain the highest weighted components \cite{Vo2019b}. 

\section{Computing the GLMB Posterior}

In this section, we summarize the Gibbs sampler proposed in \cite{Vo2019b}
to truncate the GLMB posterior. 

\subsection{Sampling Distributions\label{subsec:Stationary-distribution}}

The GLMB posterior is truncated from some discrete probability distribution
$\pi$ of association maps $\gamma_{0:k}$, so that those maps with
higher weights are more likely to be chosen. Starting with $\mathcal{L}(\gamma_{0})=\emptyset$,
we consider
\begin{align}
\pi(\gamma_{0:k})= & \ \overset{k}{\underset{j=1}{\prod}}\pi^{(j)}(\gamma_{j}|\gamma_{0:j-1})\propto w_{0:k}^{(\gamma_{0:k})},\label{eq:stationary_dist_gamma}
\end{align}
where
\begin{align}
\pi^{(j)}(\gamma_{j}|\gamma_{0:j-1})\propto & \ w_{j}^{(\gamma_{0:j})}\label{eq:factored_dist_of_gamma}\\
\propto & \ 1_{\Gamma_{j}}(\gamma_{j})1_{\mathcal{F}(\mathbb{B}_{j}\uplus\mathcal{L}(\gamma_{j-1}))}(\mathcal{L}(\gamma_{j}))\nonumber \\
\times & \ [\eta_{j|j-1}^{(\gamma_{0:j}(\cdot))}(\cdot)]^{\mathbb{B}_{j}\uplus\mathcal{L}(\gamma_{j-1})}.\nonumber 
\end{align}
The term $1_{\Gamma_{j}}(\gamma_{j})$ makes sure that $\gamma_{j}$
is in the space of all association maps at time $j$, and $1_{\mathcal{F}(\mathbb{B}_{j}\uplus\mathcal{L}(\gamma_{j-1}))}(\mathcal{L}(\gamma_{j}))$
makes sure that only values of $\gamma_{j}$ on $\mathbb{B}_{j}\uplus\mathcal{L}(\gamma_{j-1})$
need to considered. 

Given a valid $\gamma_{0:k}$, the Gibbs sampler constructs a sequence
of iterates, such that the next iterate $\gamma_{0:k}^{'}$ is generated
from $\gamma_{0:k}$ by sampling each $\gamma_{j}^{'}(\ell_{n})$
from the conditional 
\begin{align}
\pi_{j,n}(\alpha| & \overbrace{\gamma_{0:j-1}^{'}}^{\mathrm{past}},\overbrace{\gamma_{j}^{'}(\ell_{1:n-1})}^{\mathrm{current\ (processed)}},\overbrace{\gamma_{j}(\ell_{n+1:|\mathbb{L}_{j}|})}^{\mathrm{current\ (unprocessed)}},\overbrace{\gamma_{j+1:k}}^{\mathrm{future}})\nonumber \\
\propto & \ \pi(\gamma_{0:j-1}^{'},\gamma_{j}^{'}(\ell_{1:n-1}),\alpha,\gamma_{j}(\ell_{n+1:|\mathbb{L}_{j}|}),\gamma_{j+1:k})\label{eq:full_gibbs_sampling_nth_cond_dist}
\end{align}
for each $j\in\{1:k\}$, $\ell_{n}\in\{\ell_{1:|\mathbb{L}_{j}|}\}$.
Let $\gamma_{\bar{j}}\triangleq(\gamma_{0:j-1},\gamma_{j+1:k})$,
\begin{equation}
\eta_{j,n}^{(\gamma_{\bar{j}})}(\alpha)\triangleq\negthinspace\negthinspace\overset{k\lor(t(\ell_{n})+1)}{\underset{i=j}{\prod}}\negthinspace\negthinspace\eta_{j|j-1}^{(\gamma_{0:j-1}(\ell_{n}),\alpha,\gamma_{j+1:i}(\ell_{n}))}(\ell_{n}),\negthinspace\negthinspace\negthinspace\label{eq:full_gibbs_sampling_joint_weight_dist}
\end{equation}
\[
M_{\beta}^{(S)}(\alpha)\triangleq\begin{cases}
\delta_{\beta}[\alpha], & \alpha<0\\
1, & \alpha=0\\
(1-1_{S}(\alpha) & \alpha>0
\end{cases},
\]
where $a\lor b$ denotes $\text{min}\{a,b\}$. Consider $\gamma_{j}$
of the valid association history $\gamma_{0:k}$, $j\in\{1:k\}$.
Then, for any $\ell_{n}\in\mathbb{L}_{j}-(\mathbb{B}_{j}\uplus\mathcal{L}(\gamma_{j-1})),$
the conditional (\ref{eq:full_gibbs_sampling_nth_cond_dist}) is given
by,
\begin{align}
\pi_{j,n}(\gamma_{j}(\ell_{n} & )|\gamma_{j}(\ell_{\bar{n}}),\gamma_{\bar{j}})\label{eq:prop_full_gibbs_sampling_prop_dead_label_space_dist1}\\
= & \ \delta_{-1}[\gamma_{j}(\ell_{n})]\delta_{\gamma_{\mathrm{min}\{j+1,k\}}(\ell_{n})}[\gamma_{j}(\ell_{n})],\nonumber 
\end{align}
and for $\ell_{n}\in\mathbb{B}_{j}\uplus\mathcal{L}(\gamma_{j-1})$
\allowdisplaybreaks
\begin{align}
\pi_{j,n}( & \gamma_{j}(\ell_{n})|\gamma_{j}(\ell_{\bar{n}}),\gamma_{\bar{j}})\label{eq:prop_full_gibbs_sampling_prop_nth_component_dist2}\\
= & \ \eta_{j,n}^{(\gamma_{\bar{j}})}(\gamma_{j}(\ell_{n}))M_{\gamma_{\mathrm{min}\{k,j+1\}}(\ell_{n})}^{(\gamma_{j}(\ell_{\bar{n}}))}(\gamma_{j}(\ell_{n})),\nonumber 
\end{align}
and set $\gamma_{j}(\ell_{n})=-1$ for all other $\ell_{n}$. 

\subsection{Computing the Posterior}

\begin{figure}[tbh]
\includegraphics[scale=0.5]{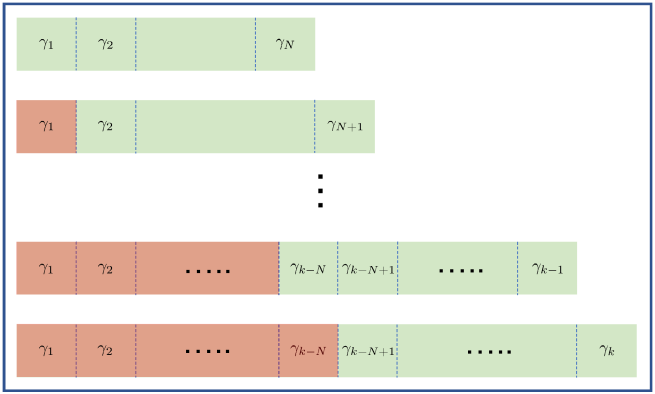}

\caption{\label{fig:Moving-window-approach}Moving window approach with window
size $N$. Maps shaded in brown are not updated.}
\end{figure}

It is clear that as the time interval, i.e., $\{0:k\}$, grows, the
dimensionality of the GLMB posterior (\ref{eq:multi_object_bayes_recursion})
increases and it becomes impractical to compute the entire posterior
at each timestep using (\ref{eq:stationary_dist_gamma}). In this
work, we propose to mitigate this computational complexity by smoothing
over fixed windows, while linking the trajectory estimates between
windows using their labels and corresponding association maps. This
approach is illustrated in Fig. \ref{fig:Moving-window-approach},
which results in a computationally efficient approximate solution
to the GLMB posterior propagation. 

Given an intial $\gamma_{0:k}$, a set of new samples can be generated
based on this technique by using Algorithm \ref{alg:Windowed-MultiScanGibbs}.
There are two methods to further improve the computational efficiency
of Algorithm \ref{alg:Windowed-MultiScanGibbs}. In the first method,
since there exist duplicate trajectories, i.e., those with the same
association history and label combination, the corresponding trajectory
and weight information can be stored and reused instead of calcuating
them each time. The second method is to paralallelize Algorithm \ref{alg:Windowed-MultiScanGibbs},
so that each new sample based on $\gamma_{0:k}$ is simultaneously
generated instead of generating one after another.

Furthermore, Algorithm \ref{alg:Windowed-MultiScanGibbs} can also
be used with a smoothing-while-filtering approach as shown in Algorithm
\ref{alg:Windowed-Smoothing-while-Filteri}. Note that the function
$\mathrm{SampleFactors}$ (in Algorithm \ref{alg:Windowed-Smoothing-while-Filteri})
is given in \cite{Vo2019b}. 
\begin{algorithm}[t]
\noindent \begin{raggedright}
inputs: $G_{0:k}=(\gamma_{0:k},w_{0:k},\tau_{0:k})$, $T$ (no. samples)
\par\end{raggedright}
\noindent \begin{raggedright}
output: $[G_{0:k}^{(t)}]_{t=1}^{T}=[(\gamma_{0:k}^{(t)}),w_{0:k}^{(t)},\tau_{0:k}^{(t)}]_{t=1}^{T}$
\par\end{raggedright}
\noindent \begin{raggedright}
\begin{tabular*}{0.8\columnwidth}{@{\extracolsep{\fill}}l}
for $t=1:T$\tabularnewline
\hspace{4mm}for $j=k-N+1:k$\tabularnewline
\hspace{4mm}\hspace{4mm}$P_{j}\coloneqq|\mathbb{B}_{j}\uplus\mathcal{L}(\gamma_{j-1})|$;
$M_{j}\coloneqq|Z_{j}|$;\tabularnewline
\hspace{4mm}\hspace{4mm}$c:=-1:M_{j}$; $\gamma'_{j}=[\ ]$;\tabularnewline
\hspace{4mm}\hspace{4mm}for $n=1:P_{j}$\tabularnewline
\hspace{4mm}\hspace{4mm}\hspace{4mm}for $\alpha=-1:M_{j}$\tabularnewline
\hspace{4mm}\hspace{4mm}\hspace{4mm}\hspace{4mm}$\varkappa(\alpha):=\pi_{j,n}(\alpha|\gamma_{j}^{'}(\ell_{1:n-1}),\gamma_{j}(\ell_{n+1:P_{j}}),\gamma_{\bar{j}});$\tabularnewline
\hspace{4mm}\hspace{4mm}\hspace{4mm}\hspace{4mm}via (\ref{eq:prop_full_gibbs_sampling_prop_nth_component_dist2})\tabularnewline
\hspace{4mm}\hspace{4mm}\hspace{4mm}end\tabularnewline
\hspace{4mm}\hspace{4mm}\hspace{4mm}$\gamma'_{j}(\ell_{n})\propto$
$\ \mathrm{Categorical}$$(c,\varkappa);$ $\gamma_{j}^{'}:=[\gamma_{j}^{'};\gamma_{j}^{'}(\ell_{n})];$\tabularnewline
\hspace{4mm}\hspace{4mm}end\tabularnewline
\hspace{4mm}\hspace{4mm}$\gamma_{j}:=\gamma_{j}^{'};$ $\gamma_{0:j}:=[\gamma_{0:j-1},\gamma_{j}];$\tabularnewline
\hspace{4mm}\hspace{4mm}compute $w_{0:k}$, $\tau_{0:k}$ via (\ref{eq:paramd_multi_scan_multi_sensor_weight_update_1})
and (\ref{eq:paramd_multi_scan_multi_sensor_trajectory_update})\tabularnewline
\hspace{4mm}end\tabularnewline
\hspace{4mm}$G_{0:k}^{(t)}:=(\gamma_{0:k},w_{0:k},\tau_{0:k});$\tabularnewline
end\tabularnewline
\end{tabular*}
\par\end{raggedright}
\caption{\label{alg:Windowed-MultiScanGibbs}Windowed-MultiScanGibbs}
\end{algorithm}
\begin{algorithm}[t]
\noindent \begin{raggedright}
input: $[G_{0:k-1}^{(h)}]_{h=1}^{H_{k-1}}$, $[T^{(h)}]_{h=1}^{H_{k-1}}$,
$T$
\par\end{raggedright}
\noindent \begin{raggedright}
output: $[G_{0:k}^{(h)}]_{h=1}^{H_{k}}$
\par\end{raggedright}
\noindent \begin{raggedright}
\begin{tabular*}{0.8\columnwidth}{@{\extracolsep{\fill}}l}
for $h=1:H_{k-1}$\tabularnewline
\hspace{4mm}$[G_{0:k}^{(h,t)}]_{t=1}^{\tilde{T}^{(h)}}\coloneqq\mathrm{Unique}(\mathrm{SampleFactors}(G_{0:k}^{(h)},T^{(h)}));$\tabularnewline
end\tabularnewline
if $k<N,$\tabularnewline
\hspace{4mm}keep $H_{k}$ best $[G_{0:k}^{(h)}]_{h=1}^{H_{k}}$\tabularnewline
\hspace{4mm}normalize weights $[w_{0:k}^{(h)}]_{h=1}^{H_{k}}$\tabularnewline
\hspace{4mm}return\tabularnewline
end\tabularnewline
keep $\bar{H}_{k}$ best $[G_{0:k}^{(h)}]_{h=1}^{\bar{H}_{k}}$\tabularnewline
for $h=1:\bar{H}_{k}$\tabularnewline
\hspace{4mm}$[G_{0:k}^{(h,t)}]_{t=1}^{T}\coloneqq\mathrm{WinMultiScanGibbs}(G_{0:k}^{(h)},T);$\tabularnewline
end\tabularnewline
$[G_{0:k}^{(t)}]_{t=1}^{\tilde{T}}\coloneqq\mathrm{Unique}([G_{0:k}^{(h,t)}]_{h,t=(1,1)}^{(\bar{H}_{k},T)});$\tabularnewline
keep $H_{k}$ best $[G_{0:k}^{(h)}]_{h=1}^{H_{k}};$\tabularnewline
normalize weights $[w_{0:k}^{(h)}]_{h=1}^{H_{k}}$\tabularnewline
\end{tabular*}
\par\end{raggedright}
\caption{\label{alg:Windowed-Smoothing-while-Filteri}Windowed-Smoothing-while-Filtering}
\end{algorithm}
\begin{figure*}[t]
\subfloat[Filtering]{\includegraphics[scale=0.31]{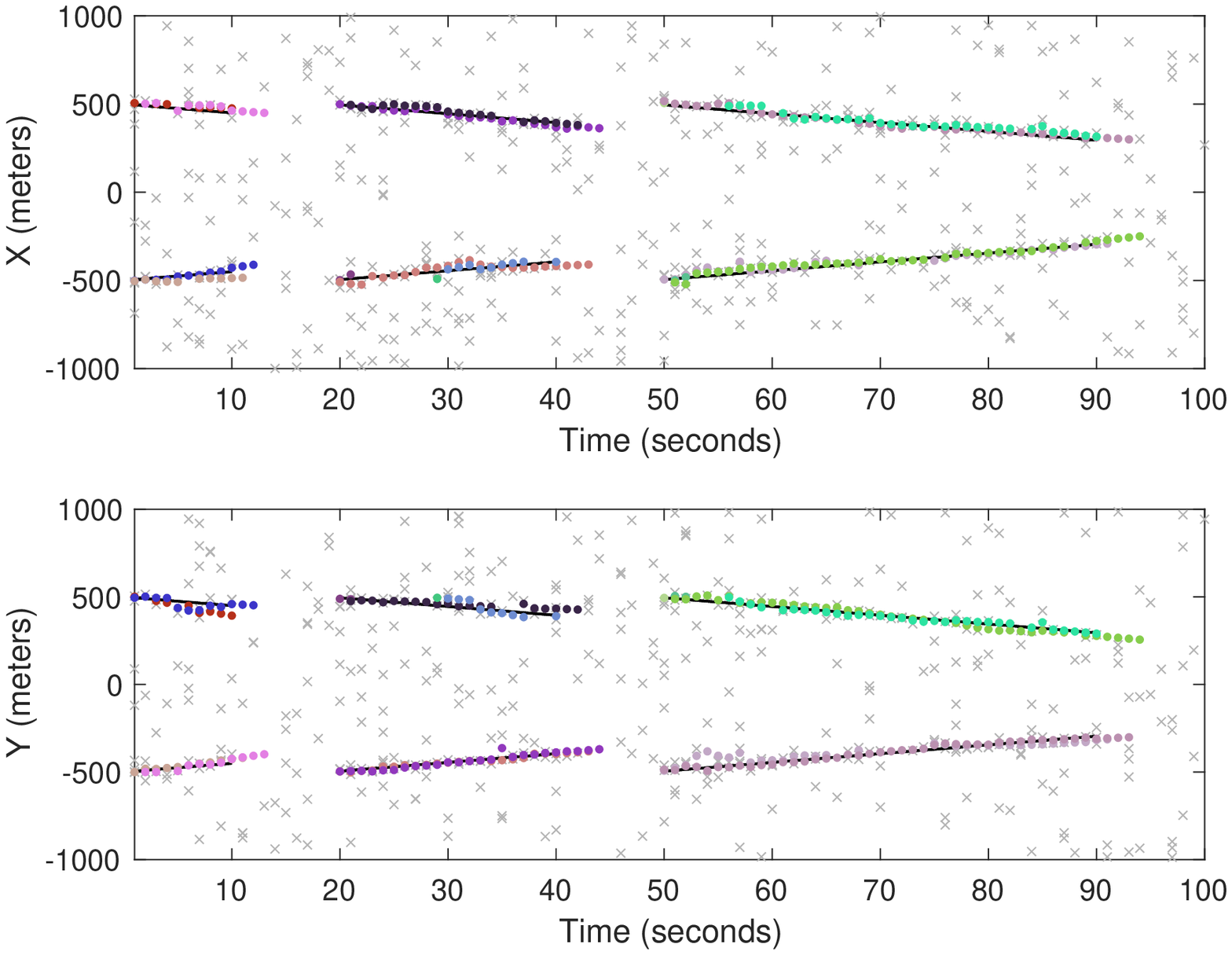}}\subfloat[Smoothing (window size = 5)]{\includegraphics[scale=0.31]{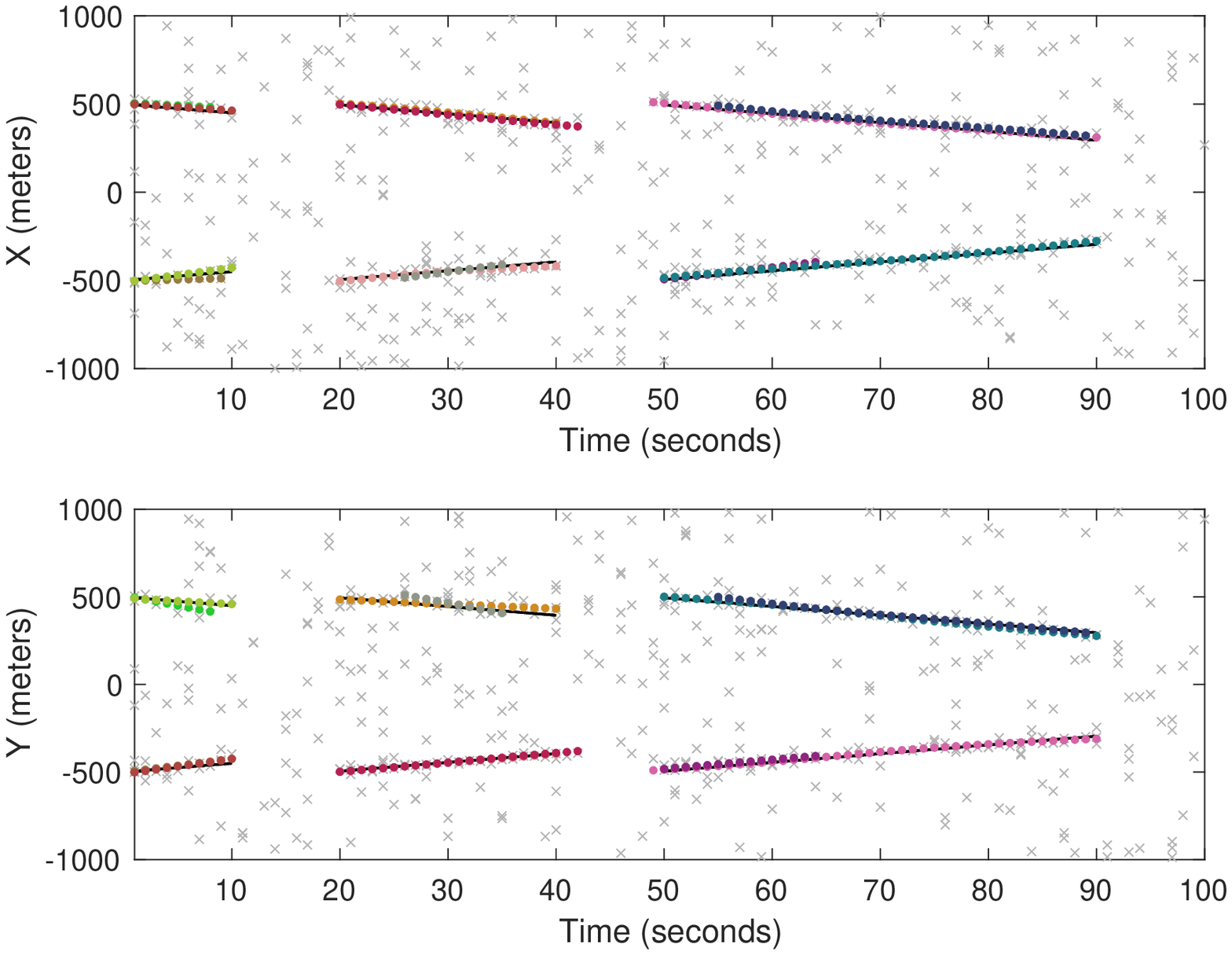}}\subfloat[Smoothing (window size = 20)]{\includegraphics[scale=0.31]{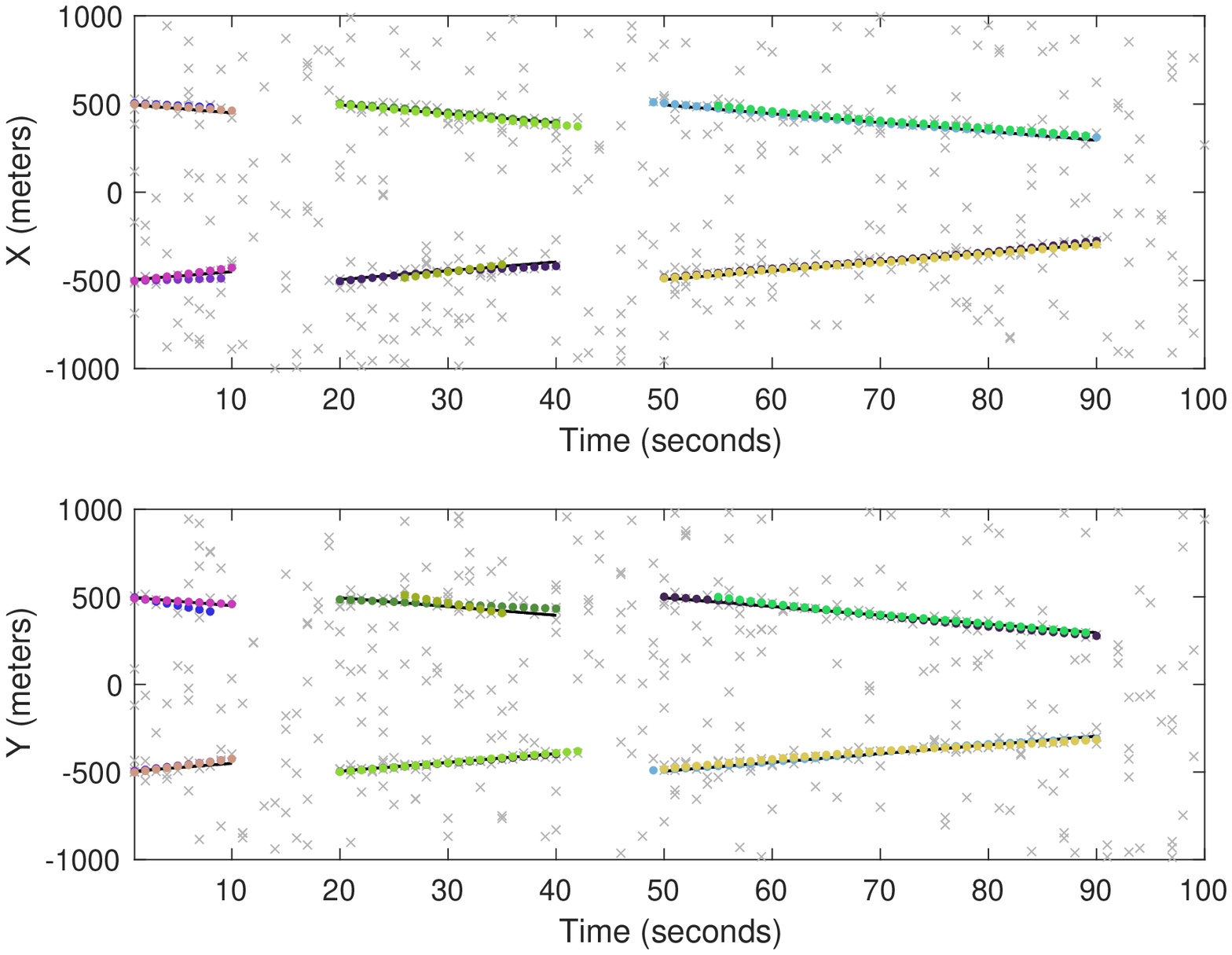}}

\caption{\label{fig:Track_Comp}Estimated trajectories (superimposed on the
ground truth - solid lines) from GLMB filtering and GLMB smoothing
with window sizes 5 and 20.}
\end{figure*}
\begin{figure*}[t]
\subfloat[Filtering]{\includegraphics[scale=0.305]{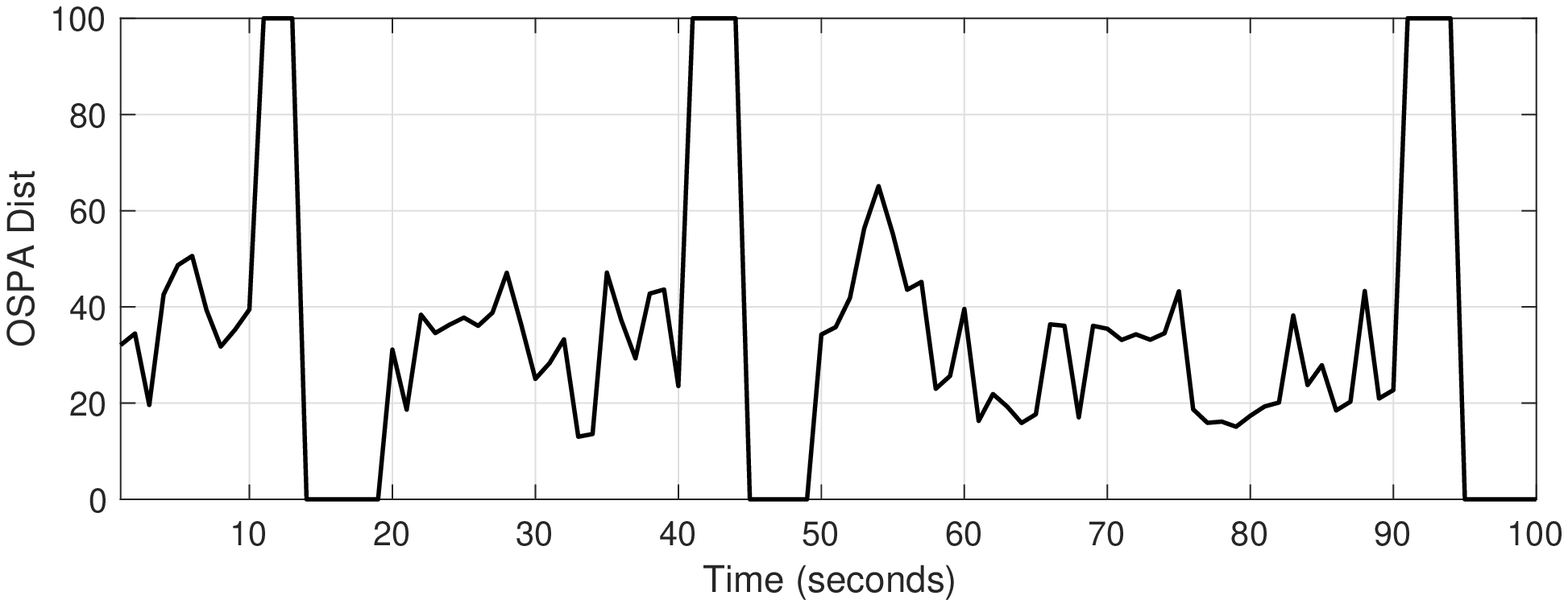}}\subfloat[Smoothing (window size = 5)]{\includegraphics[scale=0.305]{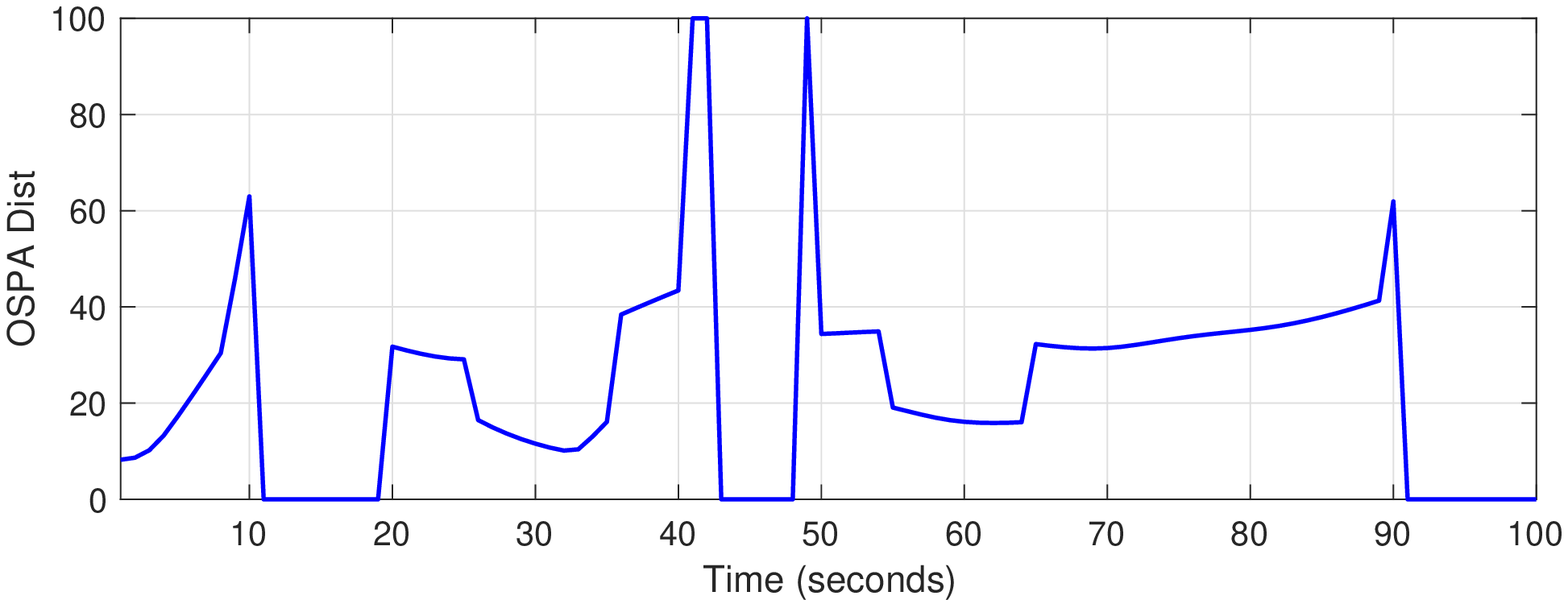}}\subfloat[Smoothing (window size = 20)]{\includegraphics[scale=0.305]{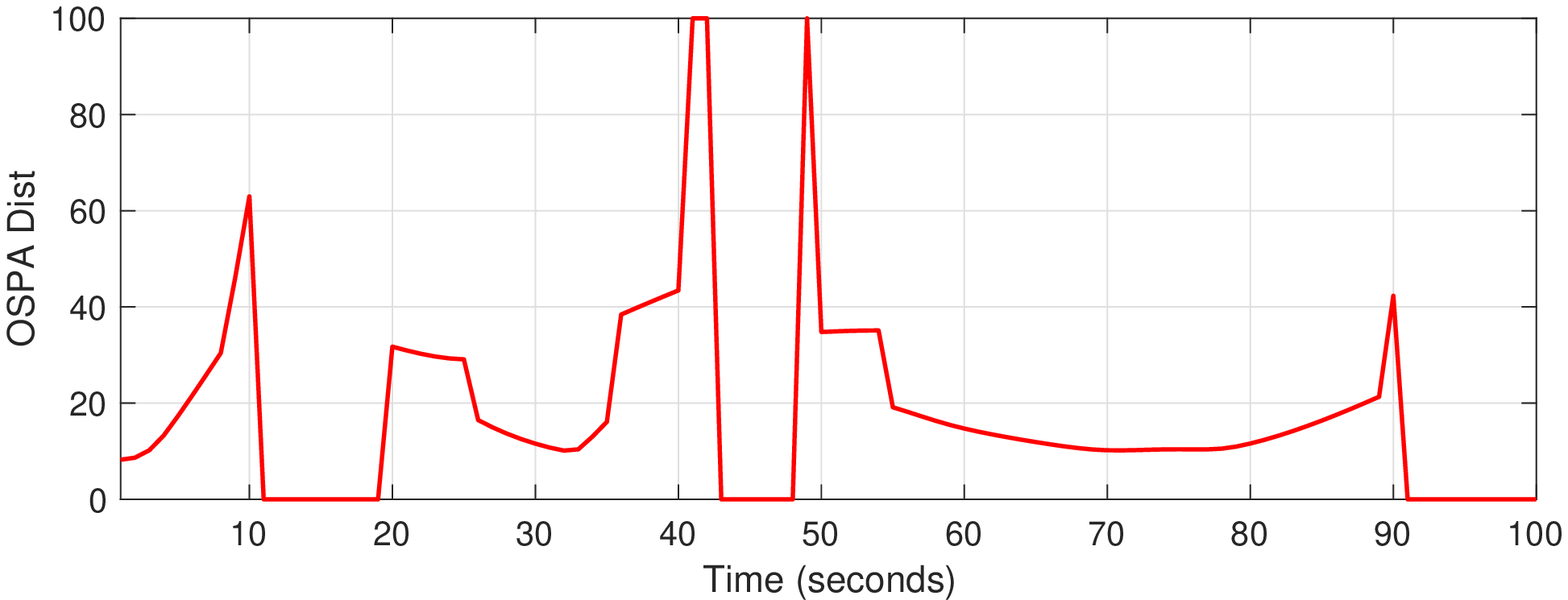}}

\caption{\label{fig:OSPA_Comp}OSPA performance from GLMB filtering and GLMB
smoothing with window sizes 5 and 20.}
\end{figure*}
\begin{figure*}[t]
\subfloat[Filtering]{\includegraphics[scale=0.305]{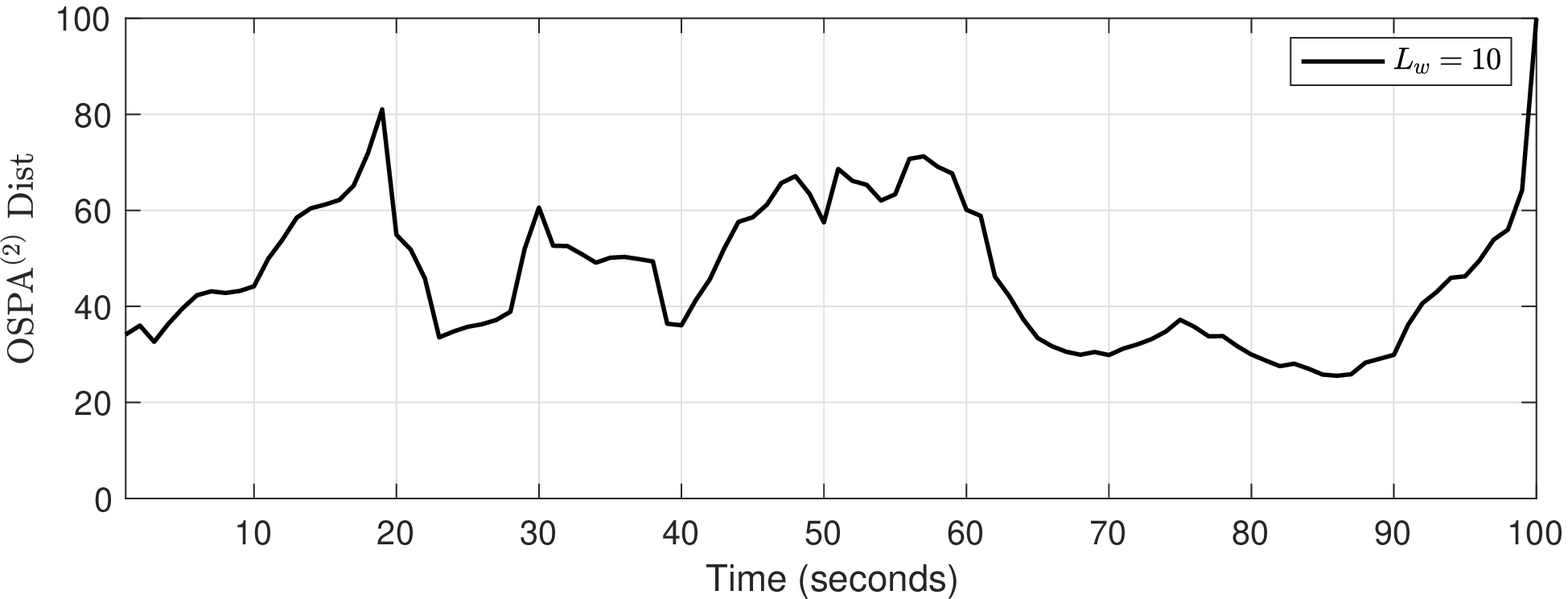}}\subfloat[Smoothing (window size = 5)]{\includegraphics[scale=0.305]{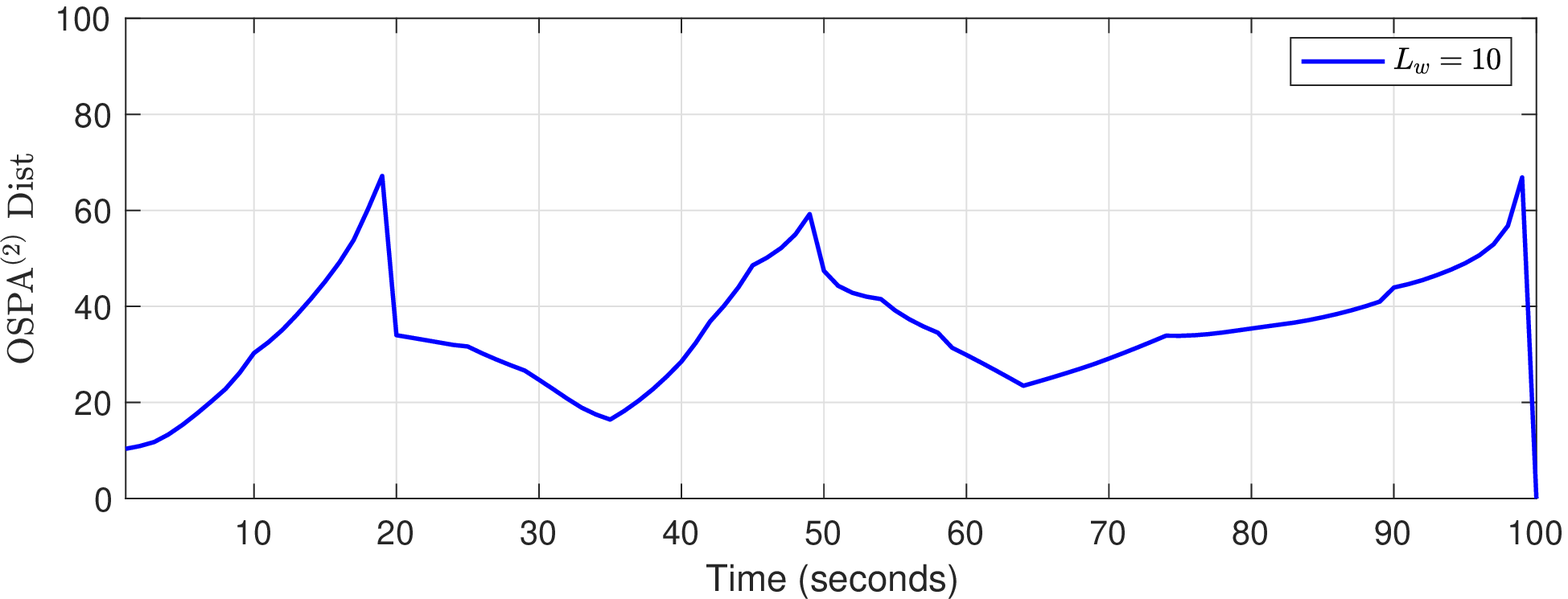}}\subfloat[Smoothing (window size = 20)]{\includegraphics[scale=0.305]{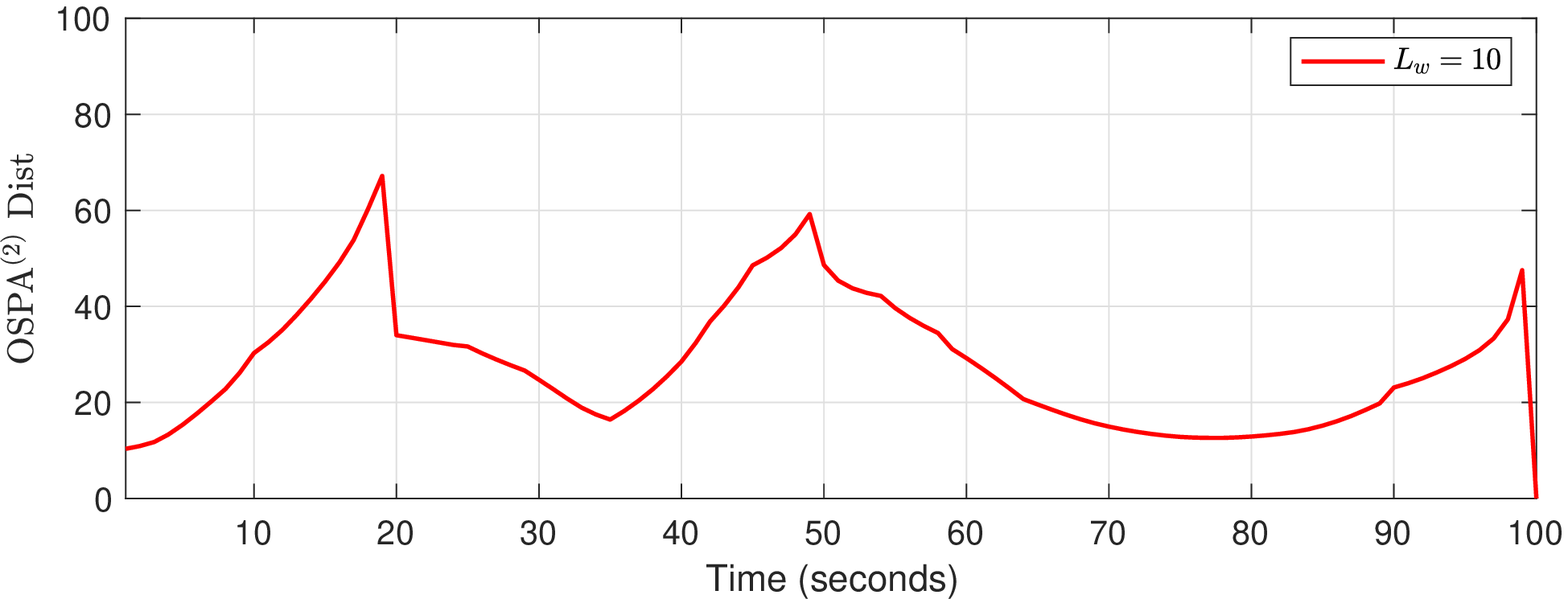}}

\caption{\label{fig:OSPA2_Comp}OSPA$^{(2)}$ (with 10 scans) performance from
GLMB filtering and GLMB smoothing with window sizes 5 and 20.}
\end{figure*}
\begin{figure*}[t]
\subfloat[Filtering]{\includegraphics[scale=0.305]{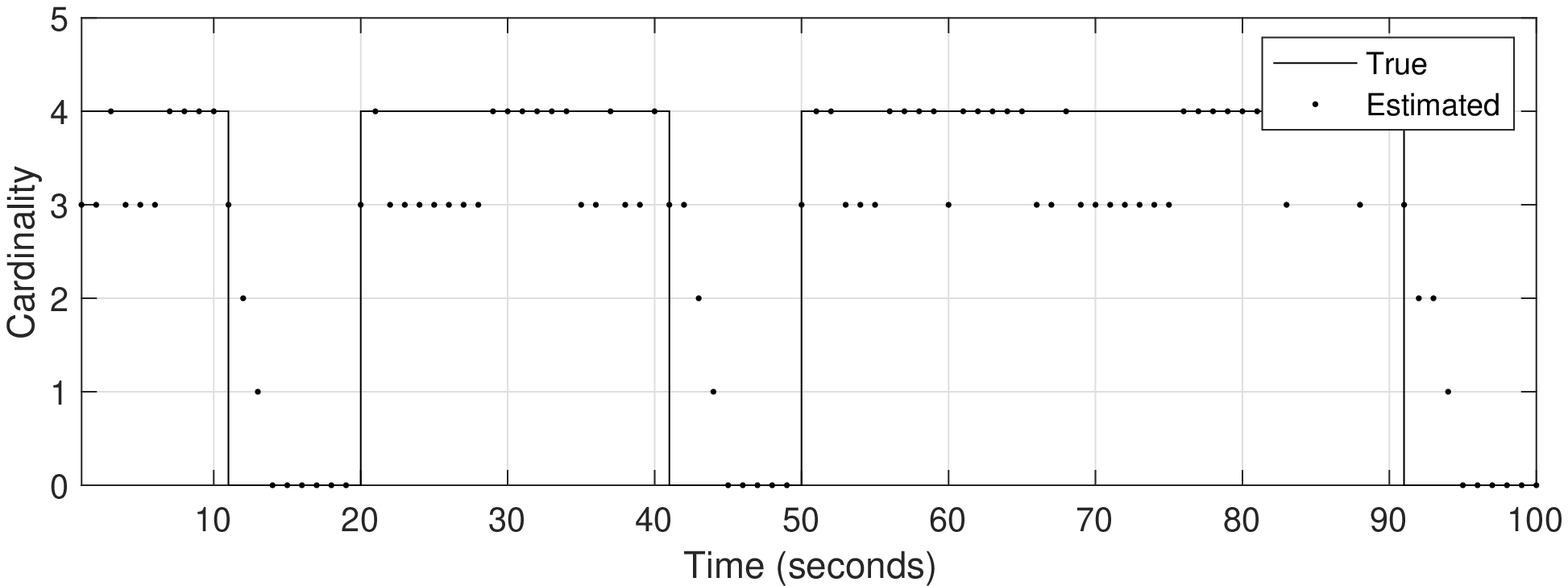}}\subfloat[Smoothing (window size = 5)]{\includegraphics[scale=0.305]{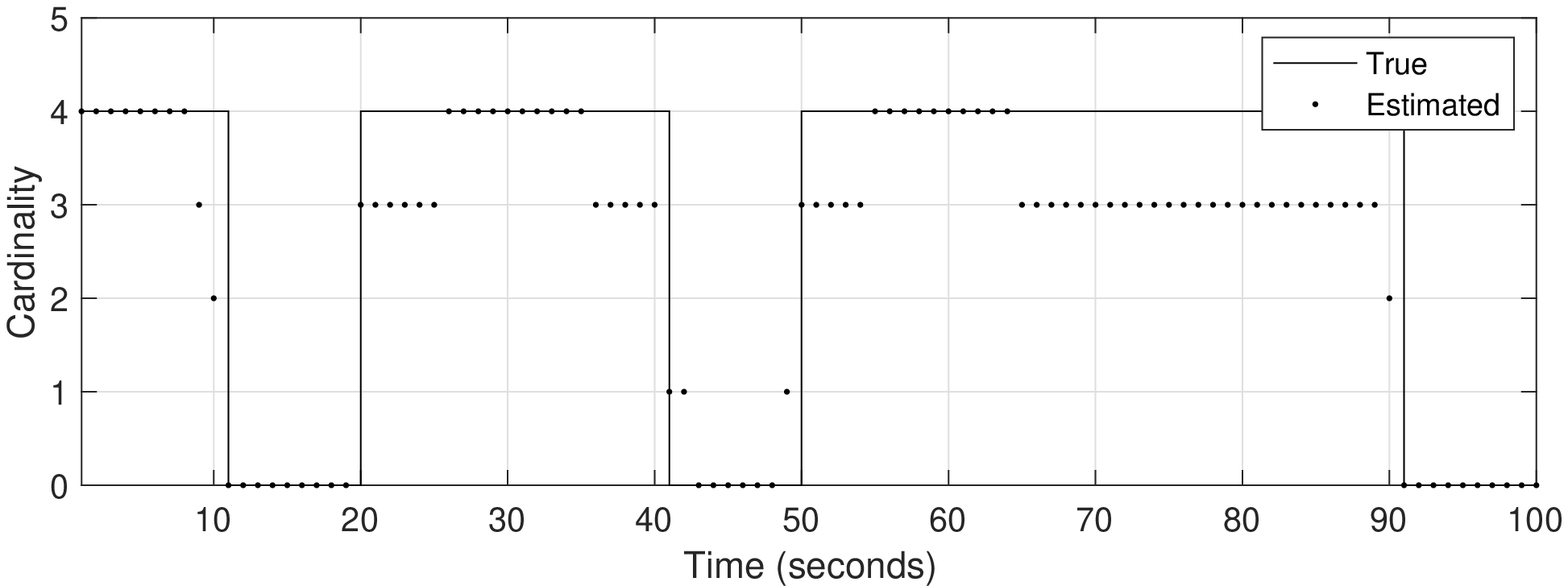}}\subfloat[Smoothing (window size = 20)]{\includegraphics[scale=0.305]{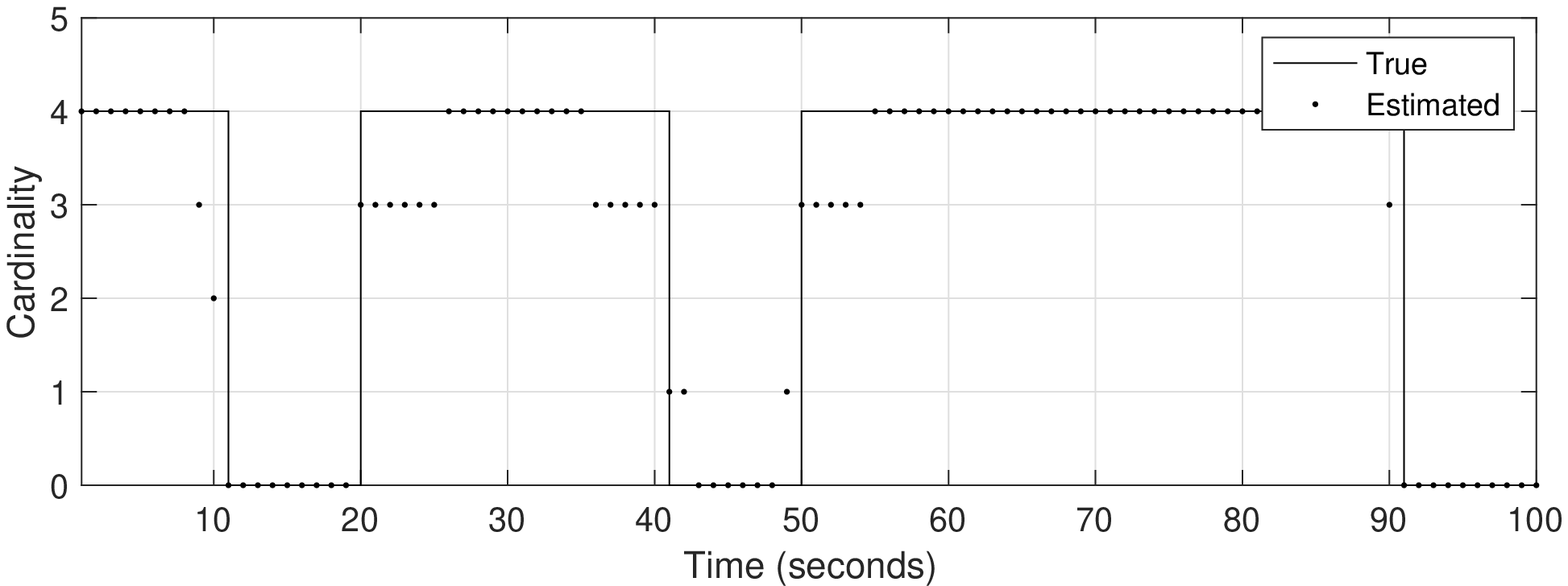}}

\caption{\label{fig:Cardinality_Comp}Comparison of estimated cardinality from
GLMB filtering and GLMB smoothing with window sizes 5 and 20.}
\end{figure*}
\begin{figure}[t]
\includegraphics[scale=0.45]{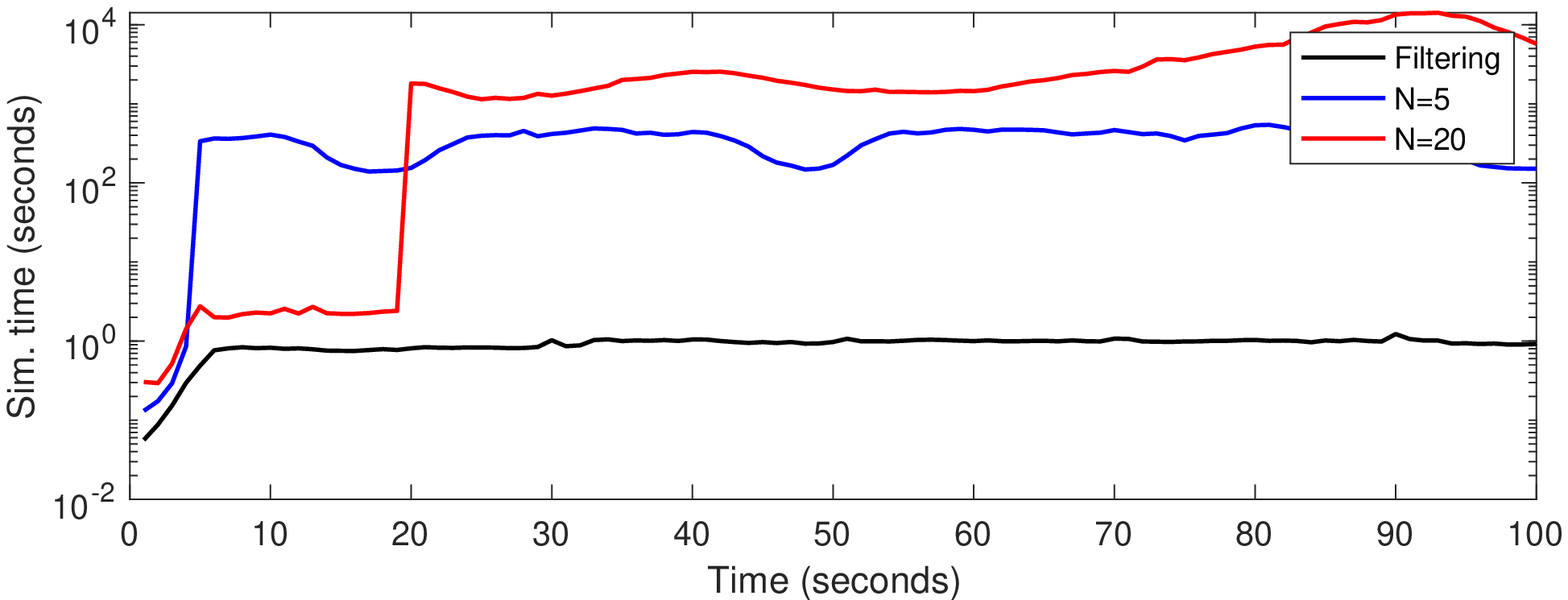}

\caption{\label{fig:Runtime_Comp}Run time comparison of GLMB filtering and
GLMB smoothing with window sizes 5 and 20.}
\end{figure}

\section{Results}

A simulation was performed to evaluate performance of the proposed
moving window based smoother. Births, deaths and movements of a set
of $12$ objects are simulated in a 2D surveillance area of $[-1000,1000]\times[-1000,1000]m^{2}$
over $100$ timesteps. The births occur at timesteps $1$, $20$ and
$50$ (respectively $4$, $4$ and $4$), and the objects born at
time $1$ die at time $10$, objects born at time $20$ die at time
$40$, and objects born at time $50$ die at time $90$. The probability
of survival of each object is set to $P_{S}(x^{(\ell)},\ell)=0.95$.
The 2D positions of the objects are measured using a sensor that adds
noise and measurement cluttter. The probability of detection of the
sensor is set to $P_{D}(x^{(\ell)},\ell)=0.3$, and clutter is modeled
as a Poisson RFS with the rate of $3$ per scan and points are uniformely
distributed over the surveillance region. The objects follow a constant
velocity motion model, and the kinematic state of an object is represented
by a 4D state vector consisting of 2D position and velocity given
by $x_{k}=[p_{x,k},\dot{p}_{x,k},p_{y,k},\dot{p}_{y,k}]$. The single
object transition density is modeled by a linear gaussian given by
$f_{S,k|k-1}(x_{k+1}^{(\ell)}|x_{k}^{(\ell)})=\mathcal{N}(x_{k+1}^{(\ell)};F_{k}x_{k}^{(\ell)},Q_{k})$
where
\[
F_{k}=I_{2}\otimes\left[\begin{array}{cc}
1 & \Delta\\
0 & 1
\end{array}\right],\hspace{0.8cm}Q_{k}=\sigma_{a}^{2}I_{2}\otimes\left[\begin{array}{cc}
\frac{\Delta^{4}}{4} & \frac{\Delta^{3}}{2}\\
\frac{\Delta^{3}}{2} & \Delta^{2}
\end{array}\right],
\]
$I_{2}$ is the $2\times2$ identity matrix, $\Delta=1s$ is the sampling
time, $\sigma_{a}=1\ m/s^{2}$, and $\otimes$ denotes the matrix
outer product. Birth objects are modeled by a Labeled Multi-Bernoulli
(LMB) Process having birth and spatial distribution parameters $\{r_{B,k}(\ell_{i}),p_{B,k}^{(i)}(\ell_{i})\}_{i=1}^{4}$,
where $\ell_{i}=(k,i)\in\mathbb{B}_{k}$, $r_{B,k}(\ell_{i})=0.03$,
$p_{B,k}^{(i)}(x^{(\ell_{i})},\ell_{i})=\mathcal{N}(x^{(\ell_{i})};m_{B,k}^{(i)},Q_{B,k})$,
$m_{B,k}^{(1)}=\ (500,0,500,0)^{T},$ $m_{B,k}^{(2)}=\ (-500,0,500,0)^{T},$
$m_{B,k}^{(3)}=\ (-500,0,-500,0)^{T},$ $m_{B,k}^{(4)}=\ (500,0,-500,0)^{T},$
and $Q_{B,k}=\ \mathrm{diag}([15,15,15,15]^{2}).$ The measurements
are modeled by a linear gaussian likelihood function of the form $g_{k}(z_{k}|x_{k}^{(\ell)})=\mathcal{N}(z_{k};H_{k}x_{k}^{(\ell)},R_{k})$,
where $R_{k}=\mathrm{diag}([30,30]^{2})$ and measurements are of
the form $z_{k}^{(v)}=[z_{x,k}^{(v)},z_{y,k}^{(v)}]^{T}$.

Algorithm \ref{alg:Windowed-Smoothing-while-Filteri} is executed
with window sizes $5$ and $20$, and compare estimated tracks, cardinality,
OSPA \cite{Schumacher2008}, OSPA$^{(2)}$ \cite{Beard18-largescale}
with the GLMB filter. Fig. \ref{fig:Track_Comp} compares the estimated
tracks. The OSPA and OSPA$^{(2)}$ matrics are compared in Fig. \ref{fig:OSPA_Comp}
and Fig. \ref{fig:OSPA2_Comp}, and it can be seen that smoothing
with window sizes 5 and 20 produce smaller OSPA and OSPA$^{(2)}$
errors than filtering, and increasing the windows size results in
smaller OSPA and OSPA$^{(2)}$ errors. The cardinality is compared
in Fig. \ref{fig:Cardinality_Comp}. The actual time (in seconds)
taken to simulate each timestep is compared in Fig. \ref{fig:Runtime_Comp}.
It is clear that the larger the window size better the performance
is, and a smaller window size with an acceptable running time can
improve the results over filtering.

\section{Conclusions}

In this paper we introduce an approximate, practical approach to the
multi-scan multi-target tracking problem using a moving window based
technique. We adopt the recent GLMB smoother based on the Gibbs sampling
based posterior truncation, and propose to recursively propagate the
multi-target posterior using the most recent $N$ scans, by linking
the measurement association maps using labels. The efficicacy of the
approach is demonstrated using a muli-target tracking simulation with
$100$ timesteps.

\section{Acknowledgement} This work was supported by the Defence Science Centre Collaborative Research Grant (in 2020).

\end{document}